\documentclass[sn-mathphys,Numbered]{sn-jnl}

\usepackage[T1,T2A]{fontenc}
\usepackage[utf8]{inputenc}
\usepackage[russian,english]{babel}
\usepackage{graphicx}%
\usepackage{multirow}%
\usepackage{amsmath,amssymb,amsfonts}%
\usepackage{amsthm}%
\usepackage{mathrsfs}%
\usepackage[title]{appendix}%
\usepackage{xcolor}%
\usepackage{textcomp}%
\usepackage{manyfoot}%
\usepackage{booktabs}%
\usepackage{algorithm}%
\usepackage{algorithmicx}%
\usepackage{algpseudocode}%
\usepackage{listings}%

\theoremstyle{thmstyleone}%
\theoremstyle{thmstyletwo}%

\theoremstyle{thmstylethree}%

\begin{document}

\title[Article Title]{Opponents and proponents of the war in Ukraine in Russian social media: who are they?}

\author[]{\fnm{Alesya} \sur{Sokolova}}\email{sokolalesja@gmail.com}
\affil[]{\orgname{Institute of Science and Technology Austria}, \city{3400 Klosterneuburg}, \country{Austria}}

\abstract{Understanding the personality of Russians who support the war in Ukraine is one of the key steps to understanding how this war became possible. However, during the war, traditional sociological methods are not always applicable. Social media provides an alternative source of what is inside people's heads. In this paper, I compare the political identities, values, and interests of social media users in Russia who hold a strong position for or against the war in Ukraine. I collect data from VK, the most popular Russian social media platform, and analyze self-filled profile information as well as the groups that the users subscribed to. I found that proponents of the war tend to have a weaker political identity (self-identified as "moderate") compared to opponents, who specify it more precisely (often, but not limited to, "liberal"). Additionally, the values of the proponents more frequently align with those promoted by the Russian government, such as orthodoxy and family. Despite these differences, pro-war and anti-war users share many common interests, as evidenced by their subscriptions to the same groups focused on music, history, and sport. When asked to state the most important trait in people (a field users can fill in VK), the most frequent answer for both groups is "kindness and honesty". The analysis results, in addition to contributing to the understanding of public opinion in Russia, can be utilized for predicting one's position on the war based on their social media profile.} 


\maketitle

\section{Introduction}\label{sec_intro}

On February 24, 2022, Russia invaded Ukraine, leading to thousands of deaths and millions of displaced in Ukraine, tightening repression and thousands of arrested dissenters in Russia, and economic and social upheaval in the rest of the world. The question of how many Russians support the invasion and why they do so became a significant part of public discussion.

The traditional methods to measure public opinion, such as polls, during times of war and under repressive pressure are skewed by respondents' fears. Experiments show that a significant percentage of Russian citizens are likely to be hiding their true views about the conflict \cite{list_exp}. Additionally, research shows that opponents of the war and the regime were more likely to be concerned about responding to pollsters and making political statements than supporters \cite{rerussia_field, chronicles2}. However, survey data from three independent projects (Levada Center \cite{levada_march}, the Chronicles project \cite{chronicles9}, and Russian Field \cite{russian_field}) indicates similar distributions of responses on crucial issues, which indicates some relevance of this survey data \cite{rerussia_field}. Such data shows that consistent war supporters and opponents make up approximately only a quarter of the population each, while the position of the remaining part tends to lean towards the support of the war, but is mixed \cite{chronicles9, russian_field}.

Another way to study public opinion and perception of the war in Ukraine is through in-depth interviews. In contrast to polls, which can capture a general trend of the attitude to the war, in-depth interviews allow tracking of precise narratives and complex perceptions, which often cannot fit into standardized sets of intelligible positions. Such a method confirms that most of the supporters of the war do not have a consistent and strong political view and, moreover, are often depoliticized \cite{imperialism_pslab, pslab_1, pslab_2}.

Both of these approaches do not usually focus on the values and personalities of respondents, which may be one of the key steps to understanding how their position developed. Meanwhile, social media provides us with an alternative source of data, from which the information on public opinion can be extracted \cite{social_media_summary, social_media_challenges}, and that was extensively used to study political phenomena \cite{social_media_polarization, social_media_fragmentation}.

In this paper, to extract the information about personalities of proponents and opponents of the war, I use data from VK, the largest Russian social media platform, which is frequently used in studies on political and social movements in Russia \cite{vk_protest_participation, vk_communities, vk_moscow_case, vk_antivaxers}. 

This social media, however, has some peculiarities that affect the contingent of its users. For the last decade, it is under the specific attention of the Centre for Combating Extremism (Centre E), a unit within the Russian Ministry of Internal Affairs heavily criticized for repressing opposition activists \cite{open_democracy_e, wired_vk, digital_repressions_review}. For publications, reposts, comments, and likes posted on their VK pages, many Russian citizens were sentenced to fines, suspended sentences, and imprisonment. For this reason, many opposition users stopped using it years before the full-scale invasion of Ukraine. However, it still remained a platform for political communication and protest coordination for some fraction of users \cite{vk_moscow_case, vk_communities}. Then, several days after the start of the full-scale invasion, Russia accepted new laws designed to criminalize public anti-war statements \cite{law_about_fakes}, pushing many of the remaining opposition users to leave VK or at least to erase there any signs of the opposition views. Additionally, Russia blocked other popular social media platforms -- Instagram and Facebook -- and announced their parent company Meta extremist organization \cite{russia_blocks}, and many pro-war users moved from there to VK.

For these reasons, VK is far from being a representative source of the statistics on the number of supporters and opponents of the war. However, in my analysis, I do not focus on the comparison of amounts of pro-war and anti-war activity. Instead, I focus on the views, values, and interests of people. While it is still possible to find enough active opponents of the war on VK, the mentioned peculiarities should not affect the reliability of the results of this study. 

For the current research, I collect a dataset of the profiles of the users who clearly express a position for or against the war in the posts they write on VK. I analyze their self-filled profile fields such as "political views", "religion", "personal priority" and "important in others". Additionally, I analyze the groups they subscribed to, to identify and compare their interests.

Since I consider only those who are confident in their position enough to write a post on it, my analysis is more focused on the consistent supporters and opponents of the war (who, according to the Chronicles project \cite{chronicles9}, make up 22\% and 20\% of the total population correspondingly), than on those whose position is not so strongly defined.

\section{Methods}\label{sec_methods}

\subsection{Dataset}

For collecting a dataset of the users who support (pro-war class) and do not support (anti-war class) the war, I used VK API. Firstly, I found pro-war and anti-war posts. Then, I added the suitable profiles of the authors of the posts to the dataset. 

To find pro-war (anti-war) posts, I used three types of search requests:

\begin{itemize}
    \item[-] Pro-war (anti-war) hashtags
    \item[-] Offensive language that supporters (opponents) of the war use with respect to opponents (supporters)
    \item[-] Terms to describe the war in Ukraine.
\end{itemize}

The detailed description of the search requests can be found in \autoref{secA1}. The date of publishing of posts was from Mai 1, 2022, to January 6, 2023.

Then, I collected the profiles of the authors of the pro-war (anti-war) posts. From this set of profiles, I removed the following:

\begin{itemize}
    \item[-] Profiles that went to both pro-war and anti-war classes (e.g. if the user uses both pro-war and anti-war hashtags in their posts)
    \item[-] Users who have less than one subscription to the groups
    \item[-] Users whose country of residence (according to the profile) is not Russia, or who did not set it
    \item[-] Users who went to anti-war class, but used pro-war hashtags that were not used for search (for the detailed description, see \autoref{secA1}).
\end{itemize}

The final dataset contains 10551 profiles, 7284 of which used pro-war vocabulary, and 3267 -- anti-war. For each profile in the dataset, the following information was extracted (the standard VK fields with a choice of answers):

\begin{itemize}
    \item[-] Sex (female or male)
    \item[-] Date of birth
    \item[-] City
    \item[-] Political views (apathetic, communist, socialist, moderate, liberal, conservative, monarchist, ultraconservative, libertarian)
    \item[-] Religion (Judaism, Orthodoxy, Catholicism, Protestantism, Islam, Buddhism, Confucianism, Secular Humanism, Pastafarianism, other)
    \item[-] Personal priority (family and children, career and money, entertainment and leisure, science and research, improving the world, personal development, beauty and art, fame and influence)
    \item[-] Important in others (intellect and creativity, kindness and honesty, health and beauty, wealth and power, courage and persistence, humor and love for life).
\end{itemize}

Also, subscriptions to the groups ("publics", as they are called in VK) were extracted for each user.

The potential problem with this dataset could be the fake accounts ("trolls"), created by companies associated with the government to promote its interests \cite{trolls}. To check if the accounts I found actually belong to real people, I compared the distribution of activity counters of the users who were allocated to the pro-war class with those allocated to the anti-war class and found that they look natural in both cases (see \autoref{secA2}). It makes it unlikely that a significant fraction of the accouts was created artificially to promote some position. This result is not surprising: the goal of trolls is to imitate public activity ("astroturfing"), and for this, writing comments and putting "likes" is more efficient than writing posts on a personal page. Additionally, trolls tend to not use public accounts (so their posts and other personal information are not accessible from search or VK API) \cite{prigozhin_liberty}.

\subsection{Considering bias in the distributions}\label{sec_correction}

To check the accuracy of the classification of posts as pro-war or anti-war based on keywords, I read 311 posts manually. From those classified as anti-war (157 posts), 95 (60\%) posts actually contained anti-war positions, 43 (27\%) contained pro-war positions, and from 19 (12\%) I could not determine the position. From those classified as pro-war (154 posts), 144 (93.5\%) posts contained pro-war positions, 2 (1.3\%) -- anti-war positions, and from 8 (5.2\%) I could not determine the position.

I assumed that if the position is undefined from the posts, then the user supports or opposes the war with equal probability. Then, the fraction of actual opponents of the war in the class of anti-war users is 67\%, and the fraction of proponents of the war in the class of pro-war users is 96\%.

To calculate the error of these values, I quadratically summed up half of the fraction of undefined posts with the statistical error of the binomial distribution. As a result, the fraction of the anti-war users in the anti-war class is

\begin{equation}
    f_{\textrm{anti-war}} = 0.67 \pm 0.07,
\end{equation}
and the fraction of pro-war users in the corresponding class is
\begin{equation}
    f_{\textrm{pro-war}} = 0.96 \pm 0.03.
\end{equation}

Knowing these values, the initial distributions can be corrected to remove the bias caused by inaccurate classification. If $\tilde{p}_{\textrm{anti-war}}(x)$ and $\tilde{p}_{\textrm{pro-war}}(x)$ are the distributions of some parameter in the biased dataset (e.g. x = \{"political views" = "liberal"\}, and $\tilde{p}_{\textrm{anti-war}}(x)$ is a fraction of the users in anti-war class who specified their political views as liberal), and $p_{\textrm{anti-war}}(x)$ and $p_{\textrm{pro-war}}(x)$ are the actual unbiased distributions, then:

\begin{equation}
    \tilde{p}_{\textrm{anti-war}}(x) = f_{\textrm{anti-war}} p_{\textrm{anti-war}}(x) + (1-f_{\textrm{anti-war}}) p_{\textrm{pro-war}}(x),
\end{equation}
\begin{equation}
    \tilde{p}_{\textrm{pro-war}}(x) = f_{\textrm{pro-war}} p_{\textrm{pro-war}}(x) + (1-f_{\textrm{pro-war}}) p_{\textrm{anti-war}}(x).
\end{equation}

From these equations, the unbiased distributions can be easily expressed in terms of dataset distributions:

\begin{equation}\label{corrected1}
    p_{\textrm{anti-war}}(x) = \frac{f_{\textrm{pro-war}} \tilde{p}_{\textrm{anti-war}}(x) - (1-f_{\textrm{anti-war}}) \tilde{p}_{\textrm{pro-war}}(x)}{f_{\textrm{anti-war}} + f_{\textrm{pro-war}} -1},
\end{equation}

\begin{equation}\label{corrected2}
    p_{\textrm{pro-war}}(x) = \frac{f_{\textrm{anti-war}} \tilde{p}_{\textrm{pro-war}}(x) - (1-f_{\textrm{pro-war}}) \tilde{p}_{\textrm{anti-war}}(x)}{f_{\textrm{anti-war}} + f_{\textrm{pro-war}} -1}.
\end{equation}

All the distributions in the \autoref{sec2} are corrected with \autoref{corrected1} and \autoref{corrected2}.

\subsection{Machine learning for predicting support of the war}\label{sec_machine_learning}

The information about the user subscriptions can be used to train a model for predicting positions about the war in Ukraine. To do this, I used the method described in \cite{facebook_likes}, where the personality traits and political positions were predicted based on the likes on Facebook. The only difference is that I use subscriptions instead of likes.

I constructed a matrix $A$, the rows of which are users, and the columns are accounts that are followed by more than 20 people from the dataset. Each element of the matrix $a_{ij}$ is equal to 0 or 1, depending on whether user $i$ is subscribed to account $j$. Users subscribed to less than two accounts were excluded from the sample. After that, some random pro-war users were excluded from the sample, so that the number of elements in pro-war and anti-war classes was the same.

For the resulting matrix, dimensionality reduction using Singular Value Decomposition (SVD) was applied, with a final dimension of 100. This reduced the matrix size from (5514, 11110) to (5514, 100). Each row of this matrix still corresponds to one user, while the columns now represent some combination of subscriptions.

Finally, the users, whose class was checked by reading the posts manually, were excluded from the training dataset. These users are to be used for the test of the accuracy of the model. The training was done with logistic regression.

\section{Results}\label{sec2}

\subsection{Demography}

I start the comparison of profiles of pro-war and anti-war users with demography.

\begin{figure}[h]
\centering\includegraphics[width=1\linewidth]{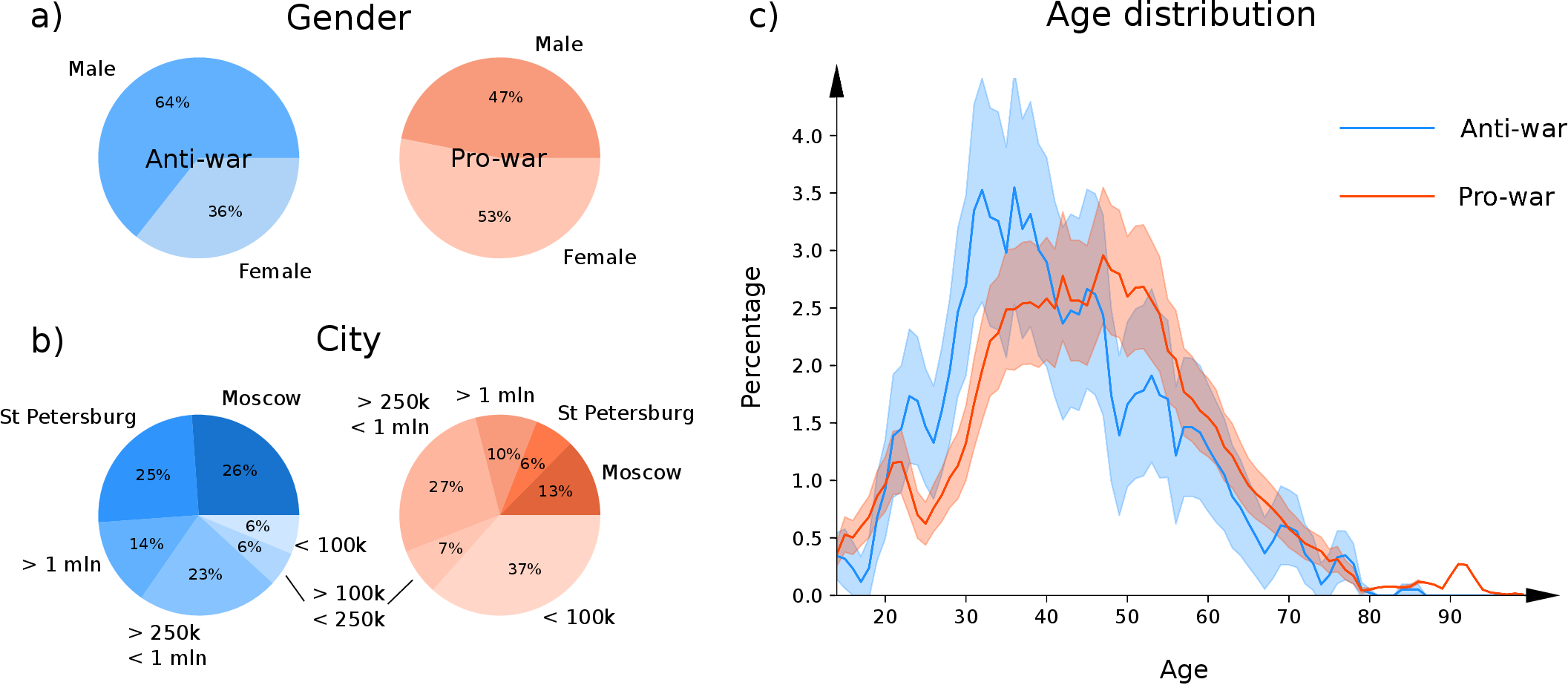}
\caption{\textbf{(a)} Distribution of users by gender, \textbf{(b)} city and \textbf{(c)} age. All the distributions are corrected as described in \autoref{sec_correction}.}\label{demography}
\end{figure}

Gender is a mandatory field in VK, so it is filled in by all the users, and its distribution in our sample is shown in \autoref{demography}(b). Among the supporters of the war, there are more women, which is surprising since most surveys \cite{russian_field, russia_watcher} indicate that women support the war less than men, and this has been a consistent result since the beginning of the Russian invasion. Therefore, it is worth talking about who writes posts about their position more often instead of who supports the war. Apparently, women write pro-war posts more actively than men.

The city is indicated by 2863 (88\%) users from the anti-war class, and 6569 (90\%) -- from the pro-war class. As shown in \autoref{demography}(b), a half of the proponents of the war in VK (51\%) lives in Moscow and Saint-Petersburg, while among pro-war users this fraction is only 18\%. In contrast, 27\% of suppporters of the war live in small localities (with population less than 100k), while only 6\% of anti-war users live in such places. It is consistent with the common beliefs that large cities are more opposition. One should notice that this result can be biased, because some of the people could leave Russia or move to the different region, but not change their location in VK.

Let's now move on to the age distribution. The date of birth is filled in by 1564 (48\%) users from the anti-war class and 3958 (54\%) -- from the pro-war. The picture \autoref{demography}(c) shows a 3 year floating average of user's age. Note that this data reflects the demographics of VK users rather than of the population as a whole, therefore their absolute values do not allow making any conclusions, and one can only compare the age difference between supporters and opponents of the war. The average age of people supporting the war is 45 years old, the median is also 45. For opponents of the war, these values are 42 and 40 years accordingly. As expected \cite{chronicles7}, the supporters of the war turned out to be older than the opponents.

\subsection{Views and values}\label{interests_sec}

The next characteristics I analyzed were the similarities and differences of self-filled user views and values between supporters and opponents of the war.

\begin{figure}[h]
\centering\includegraphics[width=0.8\linewidth]{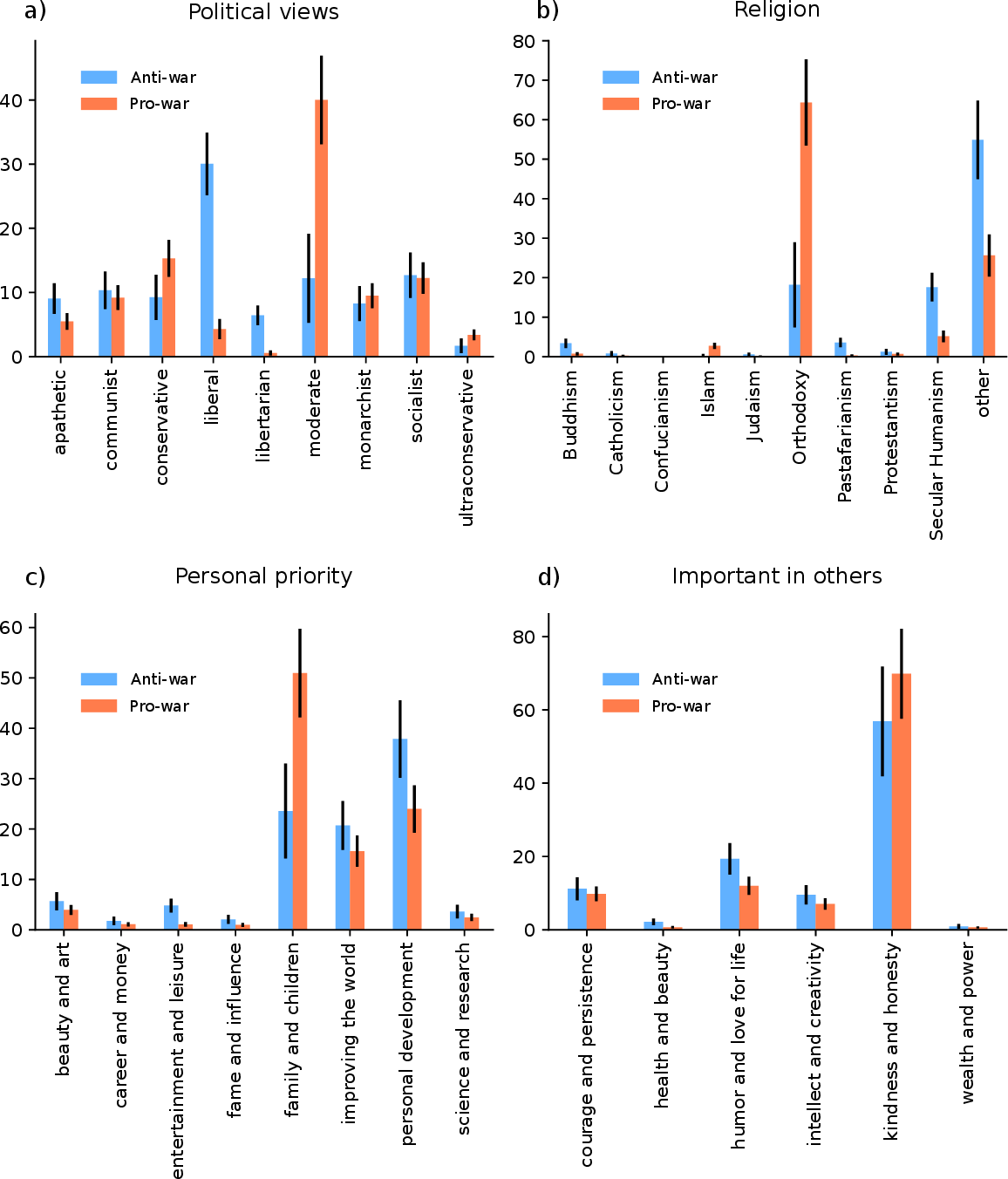}
\caption{Distribution of user views: \textbf{(a)} political views, \textbf{(b)} religion, \textbf{(c)} personal priority, \textbf{(d)} important in others. All the distributions are corrected as described in \autoref{sec_correction}.}\label{values}
\end{figure}

I start with political views (see \autoref{values}(a)) -- they are indicated by 995 (14\%) people from the pro-war class and 662 (20\%) from the anti-war. Among anti-war-minded people, the percentage of people identified as liberals is significantly larger than among pro-war, for whom it is one of the least popular political views. This result is totally consistent with the popular worldview in Russia, according to which liberal views are incompatible with pro-government ideology \cite{ria_nazism,tass_liberalism}.

Among pro-war users, the most popular views are "moderate". It is possible that this variant is a default option for people who do not have any strong political orientation. Therefore, it may indicate lower interest in politics or even political apathy of the supporters of the war. It confirms the result that Russian propaganda derives its effectiveness from political apathy rather than its ability to persuade \cite{apathy}.

Interestingly, in terms of the ratio of socialists and communists, as well as monarchists, there is almost no difference between anti-war and pro-war users. The result that the supporters of the war are fans of the USSR not more than opponents is also confirmed in \autoref{sec_inerests}.

Religious views are indicated in their profile by 482 (15\%) users with an anti-war position and 728 (10\%) with a pro-war. Only 20\% of the anti-war users define their religion as Orthodoxy, while among the supporters of the war, this number reaches almost 65\%. 

The most popular answer among the war opponents in our sample is the free-form option, which is shown as “Other” in the diagram. Here are some examples of the answers in this field (translated from Russian): “God exists”, “priest of the Orthodox Church”, “God is not exactly what the church says”, “believe in love”, “not religious”, “believe in myself ". Thus, it is difficult to determine the religiosity of this category of users.

According to surveys \cite{religiosity_levada}, 71\% of Russians consider themselves Orthodox. However, only 22\% of the respondents attended religious services more than once a year, and 49\% believe in life after death. Therefore, Orthodox identity does not necessarily mean following the Orthodox religion. Considering the result that the number of "Orthodox" people among the supporters of the war in VK is more than 3 times higher than among opponents, one can assume that Orthodox identity may be based on supporting the values and political course of the Russian Orthodox Church, which leans in line with the government course.

The next profile fields that I analyzed are "personal priority" (indicated in 589 (18\%) anti-war and 999 (14\%) pro-war users) and "important in others" (indicated in 597 (18\%) and 1066 (15\%) users, respectively). Interestingly, anti-war and pro-war users do not differ a lot in these parameters. The most significant difference is that “family and children” are more important for war supporters (about half of the pro-war users chose this option), while war opponents are slightly more likely to choose “self-development” (38\%). This difference is especially interesting in the context of the rhetoric of the Russian government and Putin, who actively advocates for "traditional values", the main component of which is a strong family \cite{trandition_values_kremlin}.

\subsection{Subscriptions and interests}\label{sec_inerests}

The next thing I explored is how the VK-publics a user follows are related to support or non-support of the war. 

\begin{figure}[h!]
\centering\includegraphics[width=1\linewidth]{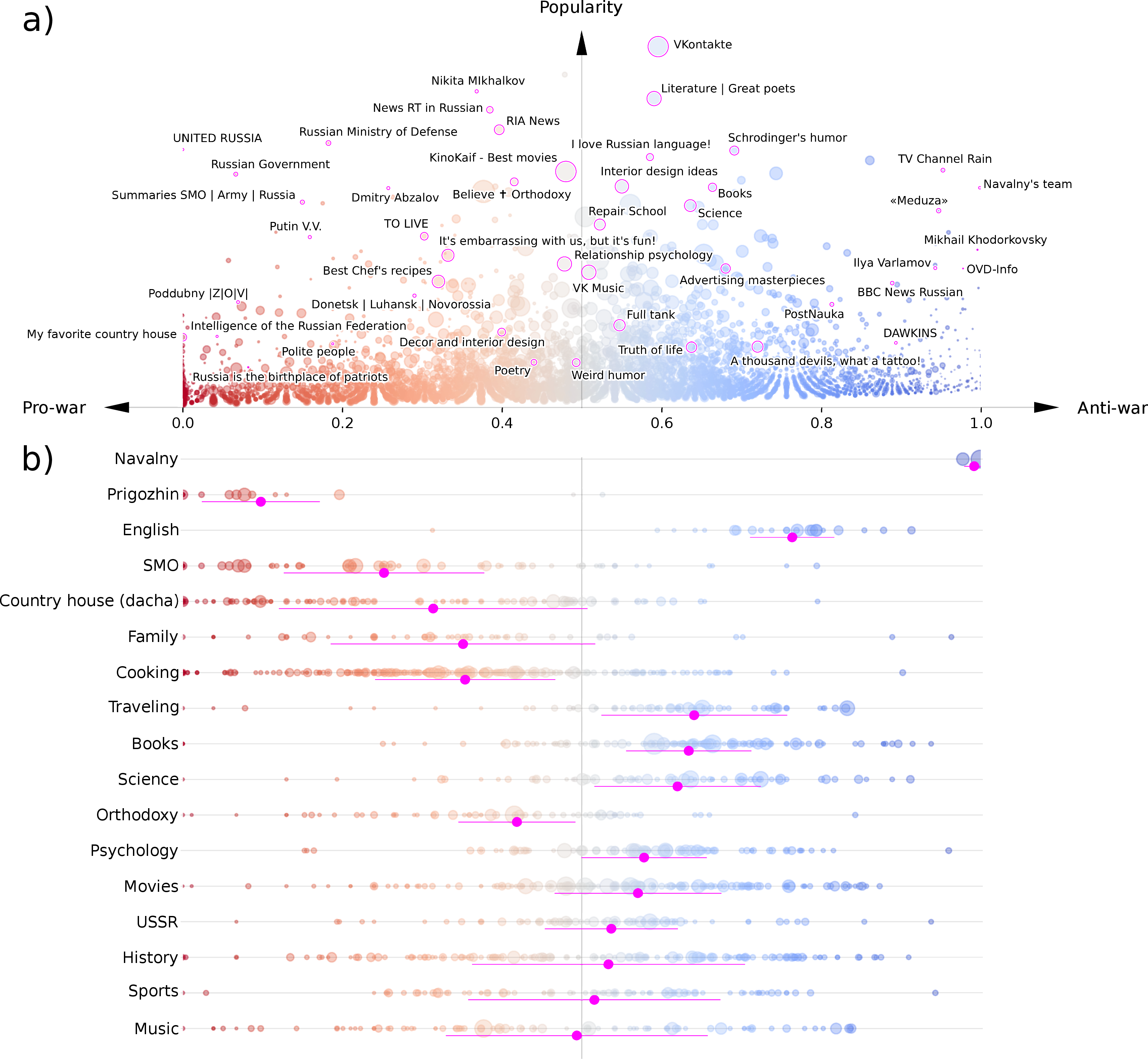}
\caption{Distribution of subscriptions of the users. \textbf{(a)} VK-publics, ranged by their popularity and a degree of pro-war/anti-war orientation. Each point is a public. The size of a point is the total number of subscribers of this public. On the vertical axis is the fraction of the users from the selection, subscribed to this public (in other words, popularity among the selection). On the horizontal axis is the fraction of anti-war users in this public (e.g. x = 0.75 means that among all the subscribers of this public from the selection, there are 75\% anti-war users and 25\% pro-war). X and Y coordinates are calculated with \autoref{publics_eq1} and \autoref{publics_eq2}. Some publics are highlighted by magenta, to give examples of their names (translated from Russian). \textbf{(b)} VK-publics classified into topics (see \autoref{sec_A_topics}), where each point corresponds to a public. The horizontal axis is the same as in (a). The size of the point here is the popularity among the selection (this parameter is on the vertical axis in (a)). The weighted average positions of the publics on the horizontal axis for each topic are shown in magenta, as well as the standard error of the distribution.}\label{publics_pic}
\end{figure}

For each public, I calculated what percentage of users from the sample are subscribed to it (i.e. popularity), which is shown on the vertical axis at \autoref{publics_pic}(a), and the fraction of the public subscribers who support and do not support the war (which is shown on the horizontal axis). 

I denote a fraction of users from the anti-war class subscribed to the public $i$ as $\tilde{p}^i_{\textrm{anti-war}} = \frac{N^{\textrm{subscribed}}_{\textrm{anti-war}}}{N^{\textrm{total}}_{\textrm{anti-war}}}$, and correspondingly $\tilde{p}^i_{\textrm{pro-war}} = \frac{N^{\textrm{subscribed}}_{\textrm{pro-war}}}{N^{\textrm{total}}_{\textrm{pro-war}}}$. These values should be corrected with \autoref{corrected1} and \autoref{corrected2} to get unbiased distributions $p^i_{\textrm{anti-war}}$ and $p^i_{\textrm{pro-war}}$. Then, these distributions are used to get $x_i$ and $y_i$ for each public $i$ in \autoref{publics_pic}(a):

\begin{equation}\label{publics_eq1}
    x_i = \frac{p^i_{\textrm{anti-war}}}{p^i_{\textrm{anti-war}} + p^i_{\textrm{pro-war}}},
\end{equation}

\begin{equation}\label{publics_eq2}
    y_i = \frac{p^i_{\textrm{anti-war}} + p^i_{\textrm{pro-war}}}{2}.
\end{equation}

The more to the right the public in the \autoref{publics_pic}(a), the more opponents of the war are subscribed to it, and the fewer supporters. And the higher it is, the more popular it is with users of both categories. Publics that are in the middle are equally popular among both supporters and opponents of the war.

All publics can be approximately divided into 3 groups. On the right are opposition accounts: "TV channel Rain", "Navalny", "Meduza", etc. On the left are pro-war political accounts: "United Russia", "Russian Ministry of Defense", "Putin V.V.", etc. And in the middle - publics on neutral topics. It is worth noting here that many (but not all) opposition publics in VK are blocked for people located in Russia, which may affect the results of the analysis.

An interesting observation here is an absence of clustering: most publics are followed by both supporters and opponents of the war. Therefore, the pro-war and anti-war users (even those who have a strong position enough to write a post on it) often coexist in the same information fields, and consume the same content, when it is not connected to the war.

\autoref{publics_pic}(b) shows interests, highlighted by keywords (\autoref{sec_A_topics}) in the names of publics, and their average position on the axis of support/non-support of the war (the same horizontal axis as in \autoref{publics_pic}(a)). For the detail of calculating the average position, see \autoref{sec_average}. The wider the magenta line, the greater the spread -- for example, among the publics about a country house, there exist the publics with almost only subscribers who support the war, but also the publics where the fraction of supporters and opponents is almost equal.

This analysis shows that many interests are shared by both anti-war and pro-war users. For example, history, sports, and music are similarly interesting for both user classes. Interestingly, almost equal number of supporters and opponents of the war are interested in USSR, which complements the result from \autoref{interests_sec}, showing that there is almost equal number of communists and socialists among pro-war and anti-war users.

However, apart from the obvious political interests, there are some topics that users of one of the groups are more interested in than the other. For example, war supporters are more interested in family and cooking, which is in line with previous results showing that family and children are more important to war supporters. At the same time, opponents of the war are more often interested in learning English and traveling.

\subsection{Predicting position based on subscriptions}

Machine learning to determine whether a person supports war or not, based on information about subscriptions. Details on how it works can be found in \autoref{sec_machine_learning}. The accuracy of the model was determined on users (with more than 2 subscribtions) whose posts were read and classified manually (72 anti-war posts and 160 pro-war ones).

It turned out that it is possible to predict that the user adheres to the anti-war position in 90\% of cases based on their subscriptions. For pro-war users, the fraction of correctly guessed positions is 70\%.

\section{Conclusion}\label{sec13}

Understanding the personality of the pro-war Russians is essential for understanding the roots of the war, as well as for finding ways to prevent similar situations in the future. Social media provide an extensive source of information on people's personalities and can be successfully utilized for this purpose.

My analysis shows that war supporters on VK are more interested in what is similar to the “traditional values” in terms of the Russian government: family and children are more important to them, and they usually consider themselves Orthodox. Meanwhile, opponents of the war are more interested in learning English, travelling, books and science.

However, despite different political views and different interests on some issues, supporters and opponents of the war have some commonalities. They constantly coexist in the same information space: they read the same content about history, sports, and music. Representatives of both groups -- both supporters and opponents of the war -- most often value kindness and honesty in people, and this value does not depend on whether the person supports the war.

About the political believes, supporters of the war most often define their political views on VK as "moderate", which may indicate their weak political identity or apoliticality, while opponents of the war often can determine their political orientation more precisely, with majority of them consisting of liberals, but also with significant fractions of socialists and conservatives. Additionally, opponents and supporters of the war are equally interested in the topic of the USSR, and the number of communists and socialists among these categories is the same.

The extra practical result of the analysis is that based on the user subscriptions, the machine learning model can predict the position regarding the war. For anti-war users, the accuracy of such prediction is 90\%, while for pro-war users it is 70\%.

\backmatter

\bmhead{Acknowledgments}

I am grateful to Novaya gazeta Europe and Teplitsa for organizing the hackathon Projector 2023, as well as my team at this hackathon, where I had a chance to present the results of the current research for the first time, improve it by gaining insights during discussions with experts and team members, and the inspiration that it gave to me. I am also thankful to Leonid Yuldashev and Equalitie for useful discussions and support.

\section*{Statements and declarations}

\bmhead{Data availability}

The anonymized data collected during the current study, as well as the code for analysis, are available in the Zenodo repository, \url{https://doi.org/10.5281/zenodo.8125674}.

\bmhead{Funding}

There was no funding for this research.

\bmhead{Competing interests}

There is no conflict of interest.

\begin{appendices}

\section{Keywords for data collection}\label{secA1}

\autoref{tab1} contains search words used for dataset collection.

\newcommand{\specialcell}[2][c]{%
  \begin{tabular}[#1]{@{}l@{}}#2\end{tabular}}

\begin{table}[h]
\caption{Search words for dataset collection}\label{tab1}%
\begin{tabular}{@{}llll@{}}
\toprule
Class & Search word -- original  & Search word -- translated & Notes\\
\midrule
Pro-war    & \#\foreignlanguage{russian}{СвоихНеБросаем}   & \#WeDontLeaveOurs  & \specialcell[t]{Meaning: "we don't leave \\ our people in trouble"} \\
Pro-war    & \#\foreignlanguage{russian}{ZaНаших}   & \#ForOurs  & \specialcell[t]{Meaning: "for our people". \\ "Z" as a military symbol \footnotemark[1]} \\
Pro-war    & ZOV   & --  & \specialcell[t]{"Z", "O", "V" as military \\ symbols \footnotemark[1]} \\
Pro-war    &  \foreignlanguage{russian}{укрофашисты}  & Ukrofascists & Meaning: "Ukrainian fascists" \\
Pro-war    &  \foreignlanguage{russian}{бандеровцы}  & Bandera & \specialcell[t]{Meaning: "people supporting \\ nationalist Stepan Bandera"} \\
Pro-war    &  \foreignlanguage{russian}{неонацисты}  & neo-nazis & -- \\
Pro-war    &  \foreignlanguage{russian}{хохлы}  & -- & \specialcell[t]{Meaning: "Ukrainians" \\(offensive)} \\
Pro-war    &  \foreignlanguage{russian}{реабилитация нацизма}  & rehabilitation of Nazism & -- \\
Pro-war    &  \foreignlanguage{russian}{\specialcell[t]{специальная военная \\ операция}}  & special military operation & \specialcell[t]{The term used by Russian \\ authorities with respect \\ to the war in Ukraine} \\
Anti-war    & \#\foreignlanguage{russian}{НетВойне}   & \#NoWar  & -- \\
Anti-war    & \#\foreignlanguage{russian}{Нет\_Войне}   & \#No\_War  & -- \\
Anti-war    & \#No\_War   & --  & -- \\
Anti-war    & \#StopRussianAggression   & --  & -- \\
Anti-war    & \#CloseTheSky   & --  & \specialcell{A call to Europe to close \\ the sky over Ukraine} \\
Anti-war    & \#RussiansAgainstWar   & --  & -- \\
Anti-war    & \foreignlanguage{russian}{Путин хуйло}   & Putin is a dick  & -- \\
Anti-war    & \foreignlanguage{russian}{zвастика}   & zwastika & Swastika with "Z" \footnotemark[1] \\
Anti-war    & \foreignlanguage{russian}{zомби}   & zombie & Zombie with "Z" \footnotemark[1] \\
Anti-war    & \foreignlanguage{russian}{zомбирование}   & zombie (process) & Zombie with "Z" \footnotemark[1] \\
Anti-war    & \foreignlanguage{russian}{путлер}   & Putler & \specialcell[t]{Combination of "Putin" \\ and "Hitler"} \\
Anti-war    & \foreignlanguage{russian}{рашисты}   & ruscists & Meaning: "Russian fascists" \\
Anti-war    & \foreignlanguage{russian}{российское вторжение}   & Russian invasion  & -- \\
\botrule
\end{tabular}
\footnotetext[1]{The letter Z is widely used as a symbol of the Russian army during the war in Ukraine. O and V are also sometimes used in this context \cite{z_nytimes}}
\end{table}

The profiles, initially classified as anti-war, but whose posts contained the hashtags from \autoref{tab2}, were removed from the dataset, since they are likely pro-war or their polarity is undefined.

\begin{table}[h]
\caption{Keywords identifying that the author of the post initially classified as anti-war should be removed from the dataset}\label{tab2}%
\begin{tabular}{@{}llll@{}}
\toprule
Keyword -- original  & Keyword -- translated & Notes\\
\midrule
\#\foreignlanguage{russian}{ZaМир}   & \#ForPeace  & "Z" as a military symbol \footnotemark[1] \\
\#\foreignlanguage{russian}{ЗаМир}   & \#ForPeace  & -- \\
\#\foreignlanguage{russian}{ЗаНаших}   & \#ForOurs  & Meaning: "for our people" \\
\#\foreignlanguage{russian}{МыВместе}   & \#WeAreTogether  & -- \\
\#\foreignlanguage{russian}{ZаПрезидента}   & \#ForPresident  & "Z" as a military symbol \footnotemark[1] \\
\#\foreignlanguage{russian}{ЗаПрезидента}   & \#ForPresident  & -- \\
\#\foreignlanguage{russian}{НетФашизму}   & \#NoFascism  & -- \\
\#\foreignlanguage{russian}{ZаРоссию}   & \#ForRussia  & "Z" as a military symbol \footnotemark[1] \\
\#\foreignlanguage{russian}{ЗаРоссию}   & \#ForRussia  & -- \\
\#\foreignlanguage{russian}{ЕР}   & \#UR  & "United Russia" (political party) \\
\#\foreignlanguage{russian}{ЕдинаяРоссия}   & \#UnitedRussia  & "United Russia" (political party) \\
\#\foreignlanguage{russian}{МариупольРусскийГород}   & \#MariupolIsRussianCity  & -- \\
\#\foreignlanguage{russian}{ДаПобеде}   & \#YesVictory  & Opposite to "no war" \\
\#\foreignlanguage{russian}{ZаПобеду}   & \#ForVictory  & "Z" as a military symbol \footnotemark[1] \\
\#\foreignlanguage{russian}{ЗаПобеду}   & \#ForVictory  & -- \\
\#\foreignlanguage{russian}{ВремяПомогать}   & \#TimeToHelp  & -- \\
\#\foreignlanguage{russian}{ГероиZ}   & \#HeroesZ  & "Z" as a military symbol \footnotemark[1] \\
\#\foreignlanguage{russian}{СВО}   & \#SMO  & Special military operation \\
\#\foreignlanguage{russian}{ДНР}   & \#DPR  & Donetsk People's Republic \\
\#\foreignlanguage{russian}{ЛНР}   & \#LPR  & Lugansk People's Republic \\
\#\foreignlanguage{russian}{ZаНамиПравда}   & \#TruthIsWithUs  & "Z" as a military symbol \footnotemark[1] \\
\#\foreignlanguage{russian}{ЗаНамиПравда}   & \#TruthIsWithUs  & -- \\
\#\foreignlanguage{russian}{ZаПравду}   & \#ForTruth  & "Z" as a military symbol \footnotemark[1] \\
\#\foreignlanguage{russian}{ZаПравду}   & \#ForTruth  & -- \\
\#\foreignlanguage{russian}{ДжекиЧан}   & \#JackieChan  & See \footnotemark[2] \\
\botrule
\end{tabular}
\footnotetext[1]{The letter Z is widely used as a symbol of the Russian army during the war in Ukraine. O and V are also sometimes used in this context \cite{z_nytimes}}
\footnotetext[2]{Hashtag \#\foreignlanguage{russian}{НетВойне} (\#NoWar) was used in a YouTube short about Jakie Chan that people reposted to VK. Its content is not connected to the war in Ukraine}
\end{table}

\section{Comparison of activity counters between classes}\label{secA2}

\begin{figure}[h]
\centering\includegraphics[width=0.8\linewidth]{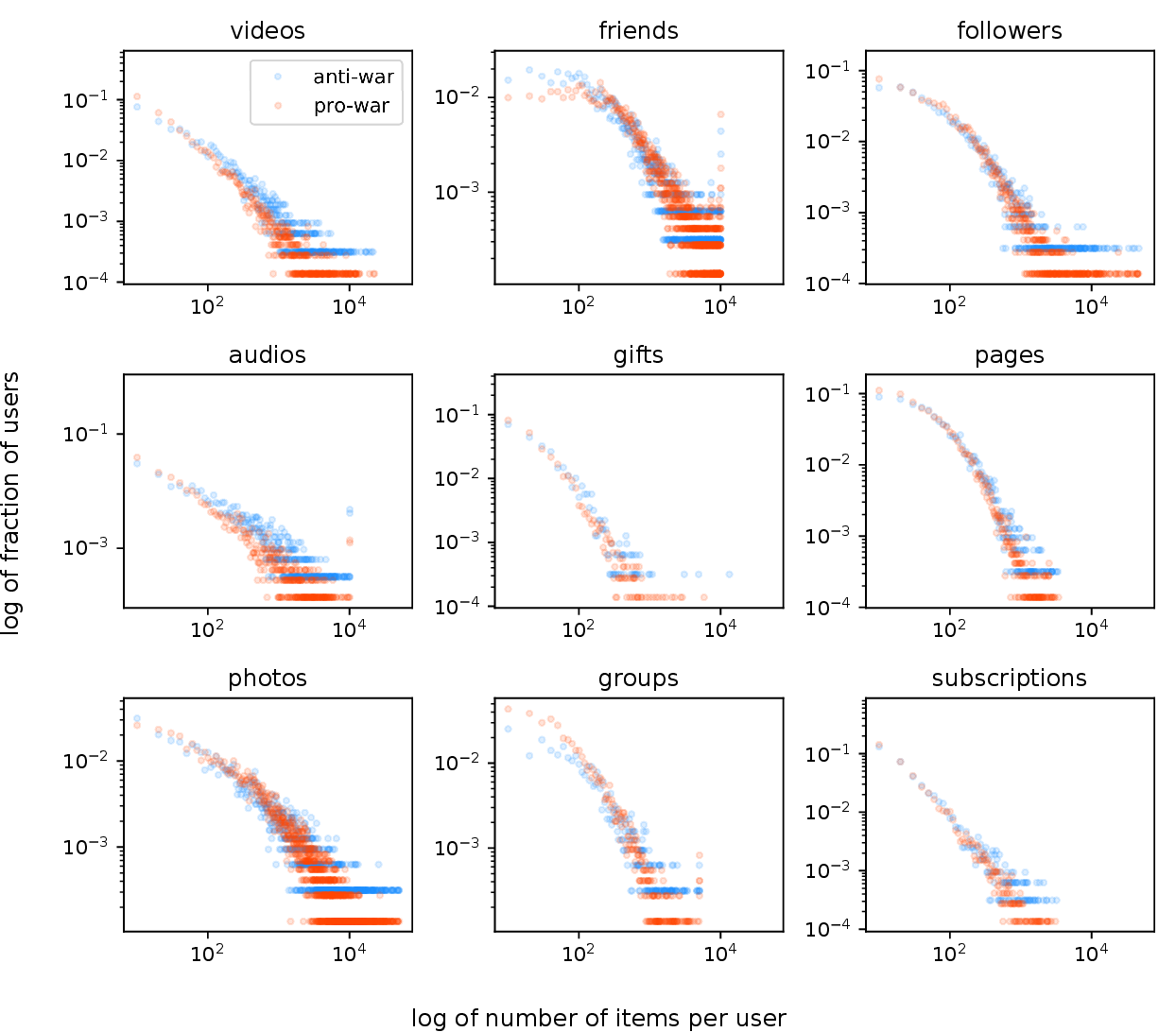}
\caption{Comparison of activity counters between classes in double-logarithmic scale}\label{counters}
\end{figure}

All the user activity counters, such as the number of videos, audio tracks, photos, friends, gifts, groups, followers, pages (meaning subscriptions to group pages, "publics"), subscriptions (to other user pages), follow the usual \cite{twitter1, twitter2, insta1} for such parameters power law, that looks linear in double-logarithmic scale for large values (\autoref{counters}). The maximum number of some of these counters in VK is restricted to 10000 (friends, audios, and groups).

Despite some differences in user counters between users from pro-war and anti-war classes, both distributions look natural, making it unlikely that a significant fraction of the accounts were created artificially specifically for promoting the pro-war position.

\section{Topics of publics}\label{sec_A_topics}

To define topics of the publics, I used keywords. The publics was considered to belong to some topics if its name contained at least one of the keywords in any form of case. E.g. the publics was considered to be about history if there was "history" or "historical" in its name. The full list of the keywords for topics is listed in \autoref{tab3}.

\begin{table}[h]
\caption{Keywords to define topics of the publics}\label{tab3}
\begin{tabular}{@{}lll@{}}
\toprule
Topic  & Keyword -- original & Keyword -- translated\\
\midrule
Navalny & \foreignlanguage{russian}{навальный, фбк}, navalny & \specialcell[t]{Navalny, \\ ACF (Anti-Corruption Foundation)} \\
Prigozhin & \foreignlanguage{russian}{пригожин, вагнер, чвк} & \specialcell[t]{Prigozhin, Wagner, \\ PMC (private military company)} \\
English & \foreignlanguage{russian}{английский}, english & English \\
SMO & \foreignlanguage{russian}{сво}, z\footnotemark[1], zov\footnotemark[1] & SMO (Special Military Operation) \\
Country house (dacha) & \foreignlanguage{russian}{дача, огород, дачный} & cottage (country house), garden \\
Family & \specialcell[t]{\foreignlanguage{russian}{семья, ребенок, родитель}, \\ \foreignlanguage{russian}{родительство}} & family, child, parent, parenting \\
Cooking & \specialcell[t]{\foreignlanguage{russian}{рецепт, кулинария, кухня}, \\ \foreignlanguage{russian}{кулинарный, вкусно, вкусный}} & \specialcell[t]{recipe, cooking, kitchen, culinary, \\ delicious} \\
Traveling & \specialcell[t]{\foreignlanguage{russian}{путешествие, туризм}, \\ \foreignlanguage{russian}{турист, путешествовать,} \\ \foreignlanguage{russian}{туристический}} & \specialcell[t]{journey, tourism, tourist, travel, \\ touristic} \\
Books & \specialcell[t]{\foreignlanguage{russian}{книга, чтение, литература}, \\ \foreignlanguage{russian}{поэзия, поэт, писатель}} & \specialcell[t]{book, reading, literature, poetry, \\ poet, writer} \\
Science & \foreignlanguage{russian}{наука, научный} & science, scientific \\
Orthodoxy & \specialcell[t]{\foreignlanguage{russian}{православие, христианство}, \\ \foreignlanguage{russian}{православный, христианский}} & \specialcell[t]{Orthodoxy, Christianity, Orthodox, \\ Christian} \\
Psychology & \foreignlanguage{russian}{психология, психологический} & psychology, psychological \\
Movies & \foreignlanguage{russian}{кино, фильм} & cinema, film, movie \\
USSR & \foreignlanguage{russian}{ссср, советский} & USSR, Soviet \\
History & \foreignlanguage{russian}{история, исторический} & history, historical \\
Sports & \specialcell[t]{\foreignlanguage{russian}{спорт, зож, тренировка}, \\ \foreignlanguage{russian}{спортивный, тренироваться}, \\ \foreignlanguage{russian}{фитнес}} & \specialcell[t]{sports, healthy, workout, sportive,\\ exercise, fitness} \\
Music & \foreignlanguage{russian}{музыка, музыкальный} & music, musical \\
\botrule
\end{tabular}
\footnotetext[1]{The letter Z is widely used as a symbol of the Russian army during the war in Ukraine. O and V are also sometimes used in this context \cite{z_nytimes}}
\end{table}

\section{Weighted average}\label{sec_average}

To calculate the average pro-war/anti-war ratio for publics on a particular topic, I use the equation for weighted average:

\begin{equation}
    \bar{x} = \frac{\Sigma x_i y_i}{\Sigma y_i},
\end{equation}
where $x_i$ and $y_i$ are calculated with \autoref{publics_eq1} and \autoref{publics_eq2} correspondingly. 

A square of the standard deviation of the weighted average is

\begin{equation}
    SE^2 = \frac{n}{(n-1) (\Sigma y_i)^2} (\Sigma (y_i x_i - \bar{y}\bar{x})^2 - 2 \bar{x} \Sigma(y_i - \bar{y})(y_i x_i - \bar{y} \bar{x}) + \bar{x}^2 \Sigma (y_i - \bar{y})^2 ),
\end{equation}
where $n$ is a number of elements (publics in the topic).

\end{appendices}

\bibliography{sn-bibliography}

\end{document}